# A tipping point for open citation data

B. Ian Hutchins, Ph.D.

Information School, University of Wisconsin-Madison, Madison, WI, USA
mailto:bihutchins@wisc.edu ORCID: 0000-0001-7657-552X

## Abstract

Open citation data can improve the transparency and robustness of scientific portfolio analysis, improve science policy decision-making, stimulate downstream commercial activity, and increase the discoverability of scientific articles. Once sparsely populated, public-domain citation databases crossed a threshold of one billion citations in February 2021, during the COVID-19 pandemic. Shortly thereafter, the threshold of one billion public-domain citations from the Crossref database alone. Since the relative advantage of withholding data in closed databases has diminished with the flood of public-domain data, this likely constitutes an irreversible change in the citation data ecosystem. The successes of this movement can guide future open data efforts.

## Main text

### Background

Science builds on the knowledge of past discoveries to advance the frontier of research. The spread of new knowledge, and the provenance of that information, is documented in the scientific citation graph. For this reason, the citations are often used in studies of knowledge flow, scientific attribution, and research assessment. Access to this information can improve discoverability, stimulate entrepreneurship, and advance basic and applied research in information networks (Dugan et al., 2017). However, for decades, the main sources of the citation data were confined by licensing restrictions from commercial providers.

The landscape has recently shifted toward open citation data. Databases like Semantic Scholar, The Lens, Microsoft Academic, and Digital Science's Dimensions have indexed citations from the literature and released them at no cost on web services or in bulk under relatively permissive licenses. However, the most permissive license is a public domain license. Importantly, two large and explicitly public-domain platforms have been developed: COCI, the OpenCitations Index of Crossref open DOI-to-DOI Citations (Peroni & Shotton, 2020), and the NIH-OCC, the National Institutes of Health Open Citation Collection (Hutchins et al., 2019). COCI indexes public citation data from Crossref (Heibi et al., 2018) that many publishers opened up in response to the Initiative for Open Citations (Dugan et al., 2017). The NIH-OCC, which I spearheaded as senior data scientist while working at NIH, is a merger of several citation databases, plus references parsed with machine learning from crawled full text articles (Hutchins et al., 2019). NIH-OCC citations are distributed through the iCite web service (iCite, 2021) and as database snapshots (Hutchins & Santangelo, 2019). These platforms have permanently released restriction-free citation data to the public, and when combined, recently hit the major milestone of **over one billion citations** in the public domain (Figure 1A). The next COCI update is expected to have this many citations from Crossref alone, following ingestion of references recently opened in Crossref by Elsevier (Plume, 2020) and the American Chemical Society (Clinton, 2021).

What fraction of citations does this represent? One estimate of the upper bound puts Elsevier's share of Crossref references at approximately 30%. Other data sources accounted for 60% of the total already, so public domain citation coverage for articles with Crossref Digital Object Identifiers may exceed 90% of the total (Dugan et al., 2017; Molteni, 2017; Waltman, 2020). Checking the Crossref application programming interface at the time of writing indicates that 88% reference-containing documents are open. Notably, the fraction when looking at journal articles only is higher, with **92%** having opened their references.

However, this accounting overestimates the fraction of all citations that are publicly available, since many print-only articles precede widespread indexing in Crossref, or publishers have not yet supplied reference information for the digital articles that are included. Here, a comparison of the biomedical literature may be instructive, since PubMed broadly indexes both digital and print-only articles. The PubMed Knowledge Graph is another source of citation information that uses Web of Science as a source of biomedical references (Xu et al., 2020), and contains references for many such historical print-only articles. Combining the references from the NIH Open Citation Collection and the PubMed Knowledge Graph shows that 86% of this union is explicitly public domain. While this estimate applies specifically to the biomedical literature, it can anchor expectations for coverage in other fields as well. Thus, the reality of citation coverage aligns closely with expectations even when accounting for print-only articles.

Importantly, in recent years, over 95% of references are explicitly public domain (Figure 1B), demonstrating nearly complete public domain coverage of recent papers. This result reinforces estimates that over 85% of Crossref citations are entering into the public domain.

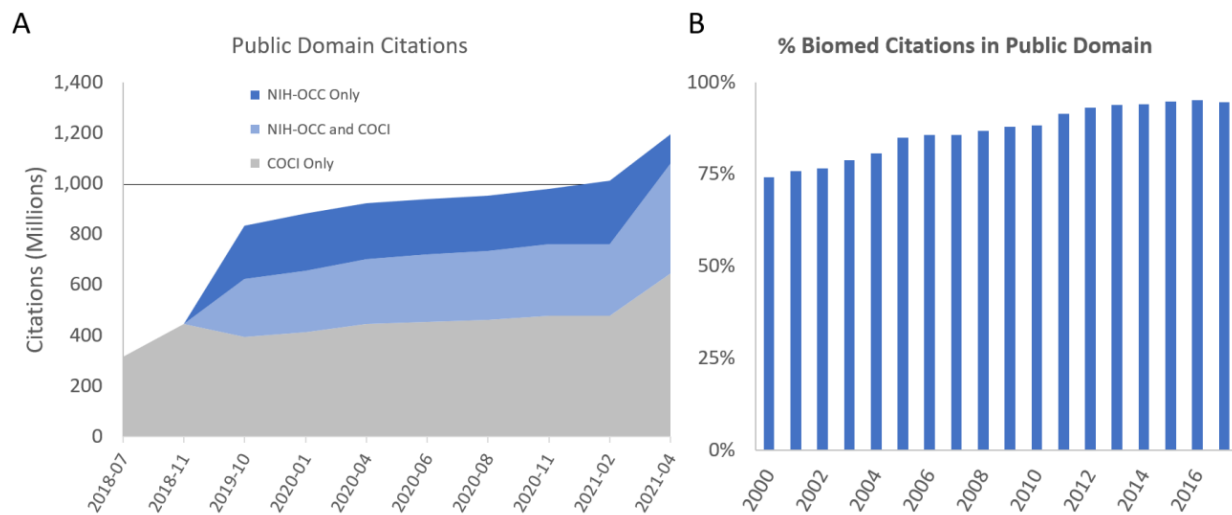

**Figure 1.** (A) Unique and overlapping citations in the public domain NIH-OCC and COCI databases. Unique public domain citations now total 1.2 billion, 46% of which are biomedical PubMed-to-PubMed citations. Black line, 1 billion citations. The final data point (2021-04) is estimated from a cache of open references from the Crossref API (the corresponding COCI dataset is still being generated as of the time of writing). (B) Fraction of biomedical citations that are in the public domain, by publication year.

## An irreversible state of affairs

When citation indexing first began, it was a laborious affair. Information extraction and processing for the Science Citation Index took teams in excess of 100 operators to generate regular database updates (Garfield, 1979). A lack of public investment necessitated that databases be paywalled in order to finance operation. Closed-access data became the norm and this persisted for decades.

Despite an explosion in the number of scientific articles published each year, the burden of indexing citations has fallen dramatically. This situation makes a prediction about the future of open citation data: that this open data ecosystem likely constitutes an irreversible change. In other words, a tipping point has been reached. It is now unlikely that the situation will revert to a closed data ecosystem, for three related reasons.

First, the increased availability of data as journals open up access to their articles (Piwowar et al., 2019), combined with these advances in hardware, open-source software, and improved algorithms, have made large-scale citation processing tractable even to small teams (Hutchins et al., 2019; Jefferson et al., 2018). Second, there are multiple, overlapping public-domain citation data providers (NIH, Crossref, and OpenCitations). These account for over 85% of citation data going forward, and ensure the data ecosystem is resilient to recissions by any one provider. For example, Microsoft announced recently that its Academic Graph will be retired at the end of 2021 ("Next Steps for Microsoft Academic – Expanding into New Horizons," 2021). Overlapping portfolios in this context is a feature, not a bug. Finally, public domain licensing guarantees that citation data is permanently available without restriction, meaning that new providers can redistribute the existing base of public domain citations even if the previous providers shut down.

The competitive benefits of closing access to citation data diminish with each new citation released to the public domain, but the benefits of open data remain. Going forward, citation data is almost completely public domain (Figure 1). In this environment, the benefits of open data overwhelm the rapidly decaying costs, even to publishers that previously monetized access to their citation data (Plume, 2020).

### Additional benefits

The scientometrics community called on publishers and funding agencies to open citation data, noting that this would improve transparency and reproducibility of scientometric analyses (Singh Chawla, 2019; Sugimoto et al., 2017). Since such analyses are used for science policy, this will improve decision-making as well. Opening citation data also improves discoverability of research articles, as readers can follow citations to more recent, relevant work.

Public domain citation data excel certain applications, compared to other permissive licenses. The first is in downstream entrepreneurship and stimulating commercial activity. Public domain citation data do not even need to be reviewed for licensing suitability, and can be immediately federated into commercial applications (Dimensions, 2019). Other licenses must be reviewed carefully, as they may contain no-commercial-use restrictions. Even some licenses that permit commercial use might exclude monetizing the redistribution of data, which could prohibit commercial curation or data cleaning. Public domain data, in contrast, are well-suited for stimulating all kinds of downstream economic activity.

A second area where public-domain licenses excel is in improving science policy. Because my professional experience lies with the US Federal Government, I will focus on that perspective, although other countries may face similar constraints. Permissive licenses targeted toward academics and research institutions may contain indemnification clauses, which render such datasets unusable in many circumstances by the US Federal Government. This is in part due to restrictions in the Antideficiency Act (O'Connell, 2014); such clauses often function as a no-government-use license. Perversely, this can drive US federal agencies to source data from closed providers, reducing the transparency and robustness of the analyses used in science policy decision-making. Public domain data are a powerful remedy for this situation.

### Conclusion

The broad coverage of public domain citation data mean that these resources are well-suited for use in scientometric analyses that traditionally used proprietary data. This community has uniquely contributed to the development of public-domain citation data, and I exhort its members to transition from using proprietary citation data to using these comprehensive public-domain data sources in their projects as much as possible. This will reduce dependence on closed data sources, increase the number of researchers using and contributing to open citation efforts, and improve the reproducibility of these projects. Additionally, I would encourage members of the community to contribute their support to other open science movements, such as the Initiative for Open Abstracts (Duce et al., 2020). The utility of open citation data will be multiplied when article text, author metadata, institutional linkages, and article classifications are also made open.

The successful movement toward public-domain citation data can offer guidance for other open-access efforts. First, identify the important gatekeepers of the current data sources. Explore win-win value propositions like discoverability and reproducibility, and clearly articulate benefits of open access. Second, build a network of open access providers. US science funding agencies aspire to data transparency, and can be powerful allies in this respect. Data acquired by US agencies under the "Unlimited Rights in Data" clauses of the Federal Acquisition Regulations and disseminated to the public can ensure their permanent open release. Finally, work toward a comprehensive release of open data. Incentives for participants in the data ecosystem will lean in favor of contributing to open data initiatives as more data are opened up and the remaining competitive value of closed data diminish. If structured well, opening data is irreversible, so the advantages to withholding data will give way while the benefits of open data remain strong.

## Acknowledgements
I thank Nees Jan van Eck for providing the Crossref bulk data for estimating the number of citations available in the 2021-05 COCI release. I also thank the NIH Office of Portfolio Analysis for providing regular iCite database snapshots containing the NIH-OCC.

## Funding statement
Support for this work was provided by the Office of the Vice Chancellor for Research and Graduate Education at the University of Wisconsin-Madison with funding from the Wisconsin Alumni Research Foundation. BIH has been supported as an NIH employee, as a trainee in the NIGMS Postdoctoral Research Associate Program, and from NIH grant F31GM080164 from NIGMS.

## Data Availability
iCite NIH Open Citation Collection data are available at Figshare (https://doi.org/10.35092/yhjc.c.4586573). The raw data supporting the analysis in Figure 1B can be found at Figshare (https://doi.org/10.6084/m9.figshare.14544405). The Crossref dataset used for the 2021-04 data point is available at Zenodo (https://doi.org/10.5281/zenodo.4748336). The PubMed Knowledge Graph data can be accessed at the University of Texas (http://er.tacc.utexas.edu/datasets/ped).

## Competing interests
BIH spearheaded the NIH-OCC as senior data scientist while working at NIH.